\documentclass{WileyMSP-template}
\usepackage{textcomp}
\usepackage{cite}

\begin{document}

\pagestyle{fancy}
\rhead{}

\title{Controlling Quantum Cascade Laser Optical Frequency Combs through Microwave Injection}

\maketitle

\author{Barbara Schneider*}
\author{Filippos Kapsalidis}
\author{Mathieu Bertrand}
\author{Matthew Singleton}
\author{Johannes Hillbrand}
\author{Mattias Beck}
\author{J\'{e}r\^{o}me Faist}

\begin{affiliations}
B. Schneider, F. Kapsalidis, Dr. M. Bertrand, M. Singleton, Dr. J. Hillbrand, Dr. M. Beck, Prof. J. Faist\\
Institute for Quantum Electronics\\
ETH-Zurich, CH-8093 Zurich, Switzerland\\
Email Address: bschnei@phys.ethz.ch

\end{affiliations}


\begin{abstract}

In this work, control over the precise state emitted by quantum cascade laser frequency combs through strong radio-frequency current modulation close to their repetition frequency is demonstrated. In particular, broadening of the spectrum from about 20 cm$^{-1}$ to 60cm$^{-1}$ can be achieved throughout most of the current dynamical range while preserving the coherence, as measured by shifted wave interference Fourier transform spectroscopy (SWIFTS). The required modulation frequency to achieve this broadening is red-shifted compared to the free-running beatnote frequency at increasing modulation powers starting from 25 dBm, whereas the range where it occurs narrows.
Outside of this maximum-bandwidth range, the spectral bandwidth of the laser output is gradually reduced and the new center frequency is red- or blue-shifted, directly dependent on the detuning of the modulation frequency. By switching between two modulation frequencies detuned symmetrically with respect to the free-running beatnote, we can generate two multiplexed spectral regions with negligible overlap from the same device at rates of at least 20 kHz. In the time-domain we show with both SWIFTS and interferometric autocorrelation (IAC) measurements a transition from quasi-continuous output to pulsed ($\tau_p \approx 55$ ps) output by ramping up the injection power to 35 dBm.

\end{abstract}

\section{Introduction}
Mid-infrared (mid-IR) spectroscopy is a powerful tool in sensing applications and environmental monitoring. Since it covers the molecular fingerprint regions of light gas molecules and functional groups of larger organic compounds, it is relevant in industrial and environmental sensing as well as biomedicine.\textsuperscript{\cite{Phillips1992BreathMedicine,Bamford2007,Sigrist2008TraceSchemes,Lee2011}} Quantum cascade lasers\textsuperscript{\cite{Faist1994}} (QCLs) are ideally suited for mid-IR spectroscopy, as they are compact and electrically pumped sources, which can operate at room-temperature while spanning the range of 3 to 20 \micro m with Watt-level output powers.\textsuperscript{\cite{Jouy2017DualPower,Schwarz2017Watt-LevelLaser/Detector}} They possess the innate ability to form optical frequency combs via four-wave mixing (FWM) caused by the strong nonlinearity of their gain-medium and waveguides\textsuperscript{\cite{Friedli2013Four-waveAmplifier,Khurgin2014CoherentLasers}} in Fabry-P\'erot and ring QCLs.\textsuperscript{\cite{Hugi2012,Burghoff2014TerahertzCombs,Piccardo2020FrequencyTurbulence,Meng2020Mid-infraredLaser}} Optical frequency combs are constituted of evenly spaced phase-locked modes.\textsuperscript{\cite{Udem2002}} The resulting high mode intensities and resolution, as compared to incoherent broadband sources, make them ideally suited for high-resolution spectroscopy, especially dual-comb spectroscopy.\textsuperscript{\cite{Coddington2016,Chen2019Mid-infraredSpectroscopy,Sterczewski2019Mid-infraredLasers,Gianella2020High-resolutionLasers}} A peculiarity of QCL frequency combs is that due to their ultrashort carrier-lifetimes on the order of picoseconds\textsuperscript{\cite{Paiella2001High-frequencyLasers}} their free-running output is not pulsed, as in most visible and near-infrared frequency combs. Rather, it is characterized by a linear frequency chirp spanning the entire round-trip period.\textsuperscript{\cite{Singleton2018EvidenceLasers,Hillbrand2020In-PhaseComb}} Therefore it is not possible to use the same characterization techniques tailored for ultrashort pulses\textsuperscript{\cite{Kane1993CharacterizationGating}} for fully explaining the underlying dynamics of QCL frequency combs. 

For optimal spectroscopic results, tunability of the spectroscopic light sources is of utmost importance. In general, QCLs allow for spectral tuning via operational temperature and electrical DC bias. In QCL frequency combs, the fast gain dynamics allow for a modulation of the carrier-populations within the cavity of at least up to tens of GHz.\textsuperscript{\cite{Martini2001High-speedLasers}} This property enables electronic read-out of the optical intermode-beating from their bias line. Likewise, external radio-frequency (RF) injection can be used as a means for injection-locking and thus manipulation of these internal oscillations.\textsuperscript{\cite{Paiella2000MonolithicLasers,Gellie2010Injection-lockingModulation,StJean2014InjectionModulation}} In earlier experiments, stabilization\textsuperscript{\cite{Hillbrand2019,Forrer2020RFBandwidth}} active mode-locking\textsuperscript{\cite{Mottaghizadeh20175-ps-longLaser,Barbieri2011CoherentSynthesis,Hillbrand2020Mode-lockedLaser}} and tuning of the free spectral range of the device\textsuperscript{\cite{Rodriguez2020TunabilityLaser}} have been demonstrated. The trade-off between RF-compatibility and thermal efficiency has been recently addressed by Kapsalidis et al.. Microstrip-waveguide geometry based designs with a double-metal waveguide, while providing superior RF-modulation capabilities, can typically only be operated at cryogenic temperatures.\textsuperscript{\cite{Unterrainer2002QuantumResonators,Calvar2013HighLine,Rodriguez2020TunabilityLaser}} This is due to the difficulty of combining a buried heterostructure and wafer bonding process. By instead employing a highly doped substrate, the optical losses could sufficiently be reduced to allow for operation close to room-temperature.\textsuperscript{\cite{Kapsalidis2021Mid-infraredGeometry}} We use this novel QCL design to demonstrate the potential of RF-modulation as a means for fast and precise tuning of the spectral and temporal properties of QCL frequency combs. Furthermore, RF-injection offers us a novel tool for probing the lasing dynamics of QCL frequency combs and thus allows us to gain further insight into the complex comb formation dynamics underlying these devices.

\section{Results and Discussion}
The device used for all measurements in this paper is a 4 mm long mid-IR QCL optical frequency comb centered at 8.2 \micro m.\textsuperscript{\cite{Kapsalidis2021Mid-infraredGeometry}} For a systematic study of the modulation-dependent behavior, the modulation frequency was swept across the free-running beatnote frequency while keeping the modulation power constant. These sweeps were repeated for different modulation powers. A detailed description of the experimental setup can be found in the experimental section below.

\begin{figure*}[htbp]
\centering
\includegraphics[width=\linewidth]{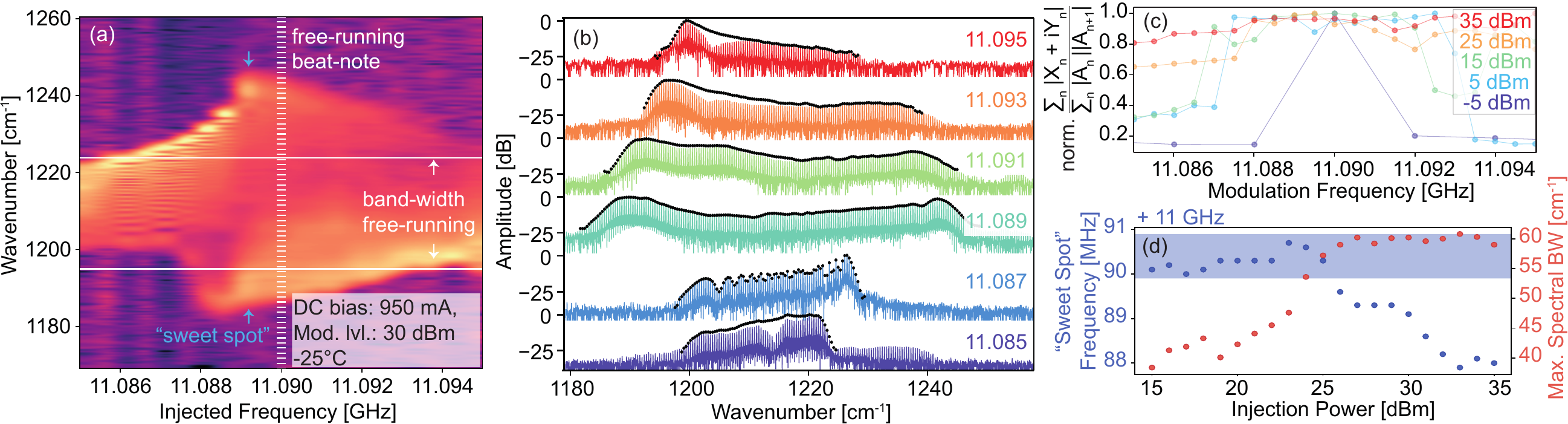}
\caption{Experimental results for modulation-frequency sweeps while injecting at constant power on the device operated with a constant DC-bias of 950 mA at -25$^\circ$C. (a) Logarithmic spectral map of the device under constant RF-injection power of 30 dBm and stepped modulation frequency between 11.085-11.095 GHz. (b) SWIFTS spectra and the corresponding spectrum products (black dots) at selected points of the spectral map in (a). (c) Ratio between the integrated spectral power from the SWIFTS spectrum and the integrated spectral power of the spectrum product indicating the average coherence for injected RF-power from -5 to 35 dBm as a function of modulation frequency. (d) Modulation frequency and bandwidth of broadest spectrum in modulation frequency spectral maps as a function of injected power.}
\label{fig:broadening1}
\end{figure*}

In \textbf{Figure \ref{fig:broadening1}} a) the resulting spectral map is shown for such a modulation frequency sweep at 30 dBm injection power. There is a clear broadening in the spectral bandwidth close to the center of the sweep with "sweet spot" marking the largest bandwidth. Notably, this point occurs at $~11.89$ GHz, about 1 MHz offset from the free-running beatnote frequency located around $11.090$ GHz. The symmetry of the QCL-spectrum is highest at the "sweet spot", as is also visible in Figure \ref{fig:broadening1} b), in which the corresponding SWIFTS spectra, $|X_n+iY_n|$, where $X_n$ and $Y_n$ are the mode amplitudes of mode $n$ of the quadrature interferograms, are plotted. At injection frequencies above this point the optical power is concentrated towards lower wavenumbers inside the spectrum and vice versa. The closest comparable behavior has been observed and modeled by Hillbrand et al.\textsuperscript{\cite{Hillbrand2020Mode-lockedLaser}} for a bifunctional active region. In their work, the spectral shape evolution has been attributed to a negative Kerr non-linearity in the system. However, they start at single-mode operation and they assume an antisymmetric shape around the maximally broadened spectrum. In our measurements, there is a clear asymmetry around the "sweet spot", which grows with increasing injection power, marked by a much faster decay in spectral bandwidth towards lower than towards higher modulation frequencies. This points to the emergence of additional factors involved in the spectral response of the QCL. As a measure of coherence, the corresponding spectrum products, $|A_n||A_{n+1}|$, where $|A_n|$ is the mode amplitude of mode $n$ of the comb measured on the MCT, are displayed as black dots for comparison with the SWIFTS spectra in Figure \ref{fig:broadening1} b). Since the SWIFTS spectra emerge from the intermode-beating of the comb, the agreement between the two datasets directly indicates the degree of mutual coherence among adjacent modes. Around the sweet spot frequency at 11.089 GHz the shapes of the SWIFTS spectrum and the spectrum product are very close throughout the entire spectral region, whereas further away there are clear gaps between the two traces when going from the spectral maximum to the center frequency. As a measure for the average coherence of the spectrum, the ratio between the integrated spectral power of the SWIFTS spectrum and the spectrum product is plotted in Figure \ref{fig:broadening1} c), as a function of modulation frequency at different injection levels. For -5 dBm injection power there is only a narrow injection range which results in high coherence. Going to 5 and 15 dBm injection, a locking range is still apparent in a step in the coherence on both ends of the sweep, however, there seems to be partial locking even outside the locking-range as indicated by the slightly increased coherence toward the edges of the sweep. Surprisingly, at these intermediate injection powers, the stable locking range stays nearly identical. Going to 25 and 35 dBm, the coherence at the edges of the sweep steadily increases, but still shows a drop compared to the coherence close to the free-running beatnote, which agrees with the results in Figure \ref{fig:broadening1} b). The extracted broadening and position of the "sweet spot" in dependence of the injection power are displayed in Figure \ref{fig:broadening1} d) for 1 dBm steps in modulation power. Interestingly, the maximum spectral bandwidth occurs around the free-running beatnote frequency at RF-powers up to 25 dBm, accompanied by a steadily growing spectral bandwidth. Above 25 dBm the maximal spectral bandwidth is clamped to around 60 cm$^{-1}$, whereas the injection frequency at which this broadening occurs starts decreasing up to an offset of 2 MHz. This indicates a fixed upper bandwidth limit of the device. The shift in frequency could be caused by an increase in the effective temperature of the active region at high-power modulation, by a Kerr-effect induced change in the effective index due to the increasing modulation depth and thus instantaneous optical power inside the cavity or a combination thereof.

\begin{figure}[htbp]
\centering
\includegraphics[width=0.45\linewidth]{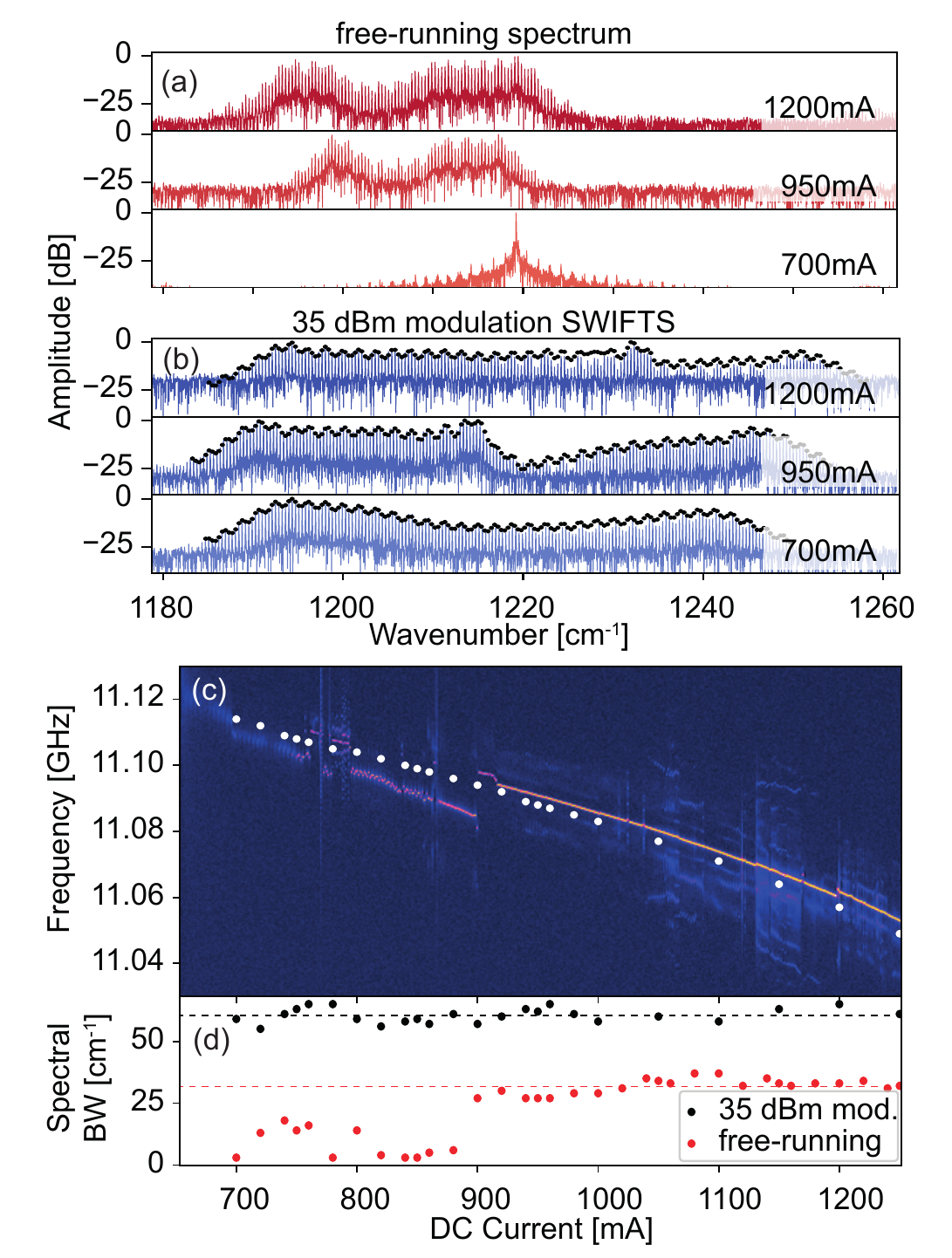}
\caption{(a) Free-running spectra of the QCl at 700, 950, 1200 mA and in (b) the corresponding SWIFTS spectra under 35 dBm injection at the "sweet spot" (maximum bandwidth) together with the spectrum product (black dots). (c) Electronic beatnote-map of the device without modulation and the injection frequencies corresponding to the largest spectral broadening at 35 dBm injection (white dots) and (d) the maximum spectral bandwidth in free-running (red) vs. under injection (black). The dashed lines represent the average bandwidth of each dataset. In the free-running case only the data starting at 900 mA bias are used.}
\label{fig:broadening2}
\end{figure}

As a further step, to study the effect of DC bias on the spectral response to RF-injection, we conducted a series of frequency sweeps at a fixed injection power of 35 dBm while stepping the DC-bias throughout the dynamical range of the QCL. For the selected bias points 700, 950 and 1200 mA, \textbf{Figure \ref{fig:broadening2}} a) shows the free-running spectra and b) the corresponding maximum bandwidth SWIFTS spectra. The free-running spectrum is single-mode at 700 mA, broadening to about 30 cm$^{-1}$ bandwidth in the range of 900 to 1200 mA. Conversely, the injected QCL spectra all cover approximately the same spectral range with bandwidths around 60 cm$^{-1}$. The coherence is close to uniform across all three characterized bias points. In Figure \ref{fig:broadening2} c), the injection frequencies at the respective "sweet spots" are plotted as white dots against the free-running beatnote map of the laser. Above 900 mA, the broadest spectra are generated 2 MHz below the free-running beatnote frequency for all bias points. The maximal spectral bandwidths at 35 dBm injection are shown in Figure \ref{fig:broadening2} d) together with the corresponding free-running bandwidths as a function of DC-bias. The injected bandwidth stays at 60 cm$^{-1}$, independent of the DC driving current.

\begin{figure}[htbp]
\centering
\includegraphics[width=0.5\linewidth]{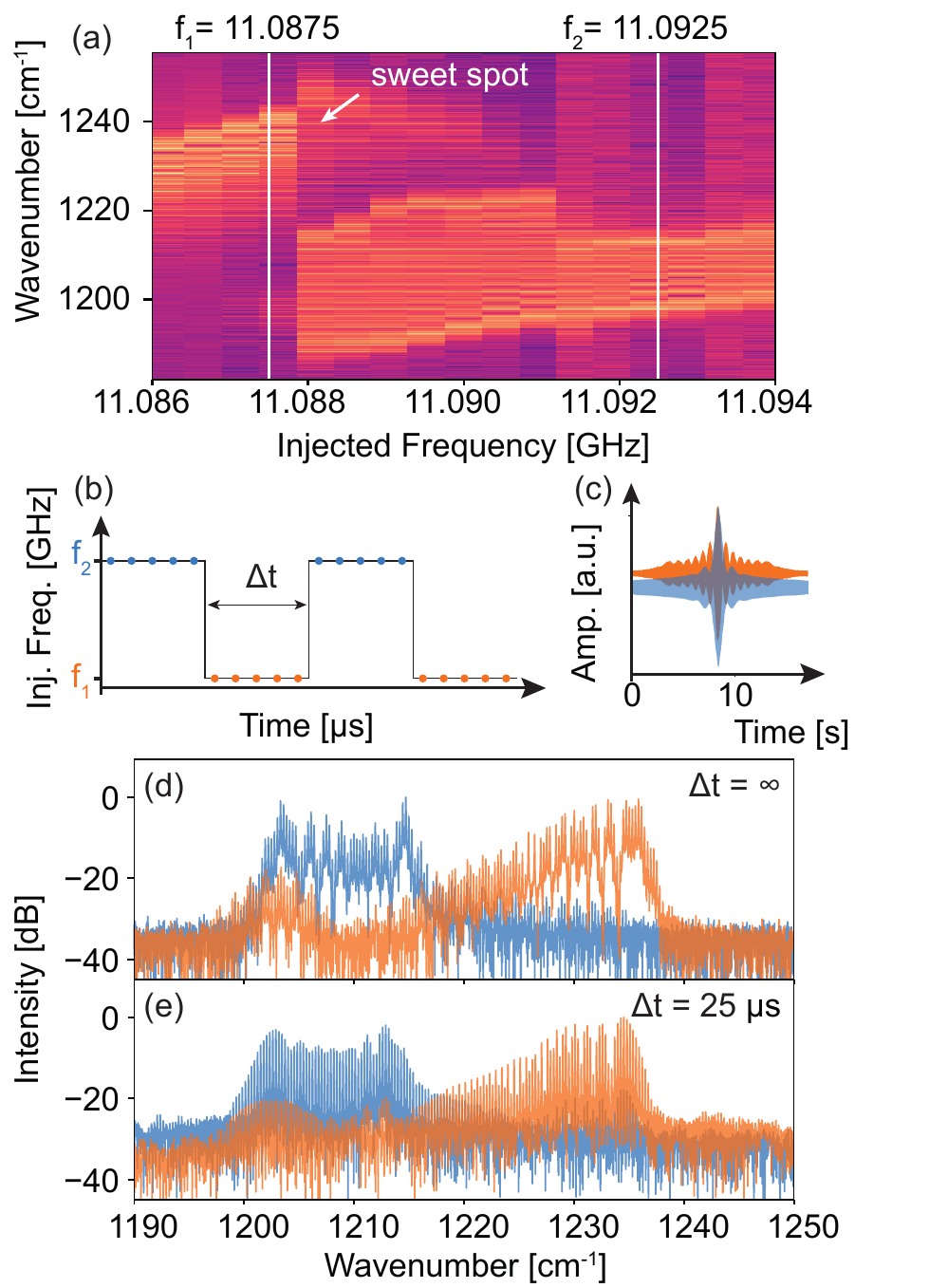}
\caption{(a) Logarithmic spectral map as a function of modulation frequency at 35 dBm injection level. The white vertical lines indicate the modulation frequencies $f_1$ and $f_2$ which generate spectra with minimal overlap. (b) Schematic depiction of the time-dependent switching between modulation frequencies $f_1$ (orange) and $f_2$ (blue) for demonstrating the acquisition of multiplexed interferograms. The dots represent the samples acquired by the oscilloscope. (c) Zoomed-in section of the experimental interferogram. The data points corresponding to the different modulation frequencies are colored according to (b). (d) Comparison of the spectra on the white lines in (a) and (e) spectra of two interferograms extracted by demultiplexing from the data displayed in (c).}
\label{fig:multiplexing}
\end{figure}

The possibility of fast, modulation frequency controlled spectral tuning opens up the opportunity for time-resolved multi-species gas monitoring by multiplexing at least two separate spectral regions, as was previously demonstrated on two-section dual-wavelength devices.\textsuperscript{\cite{Jagerska2015SimultaneousSpectroscopy, Kapsalidis2018Dual-wavelengthSpectroscopy}} In our case the device is single section and offers superior fine-tuning capabilities as we can freely switch between any spectral distribution found in the spectral map in \textbf{Figure \ref{fig:multiplexing}} a). While each individual spectrum covers a relatively small bandwidth, multiplexing several well-known spectral regions allows for monitoring multiple species based on relative intensity without the need for a spectrometer which increases the overall detection speed. At 35 dBm injection, the position of the "sweet spot" in our device is well defined with a sharp transition between a blue-shifted center-frequency of the QCL-spectrum at lower injection frequencies and a red-shifted center-frequency at higher injection frequencies, as depicted in the spectral map in Figure \ref{fig:multiplexing} a). The respective shifts are sufficiently large that there are spectra on both sides of the "sweet spot" with small overlap with respect to each other, such as the ones marked by the white lines. To experimentally verify the viability of fast spectral tuning of our QCL comb, we acquired an interferogram of the injected QCL while switching between the frequencies $f_1$ (orange) and $f_2$ (blue) (Figure \ref{fig:multiplexing} b)). A zoomed-in version of the experimentally acquired data is shown in Figure \ref{fig:multiplexing} c), where the data points are colored according to the corresponding injected frequencies $f_1$ (orange) and $f_2$ (blue). For the spectral analysis, the acquired data was demultiplexed to yield several independent interferograms with only one sample point per switching period, each. Our switching-rate and thus the achievable time-resolution was limited by the frequency-modulation capabilities of our RF-generator output, which went up to 20 kHz in square-wave operation. Comparison between the continuously modulated case and switching modulation at 20 kHz is shown in Figures \ref{fig:multiplexing} d) and e). Here, the spectrum generated under injection at $f_1$ is plotted in orange together with the spectrum at $f_2$ in blue for each case. While there is some residual overlap between the two spectral ranges, the relative intensity differs by at least 20 dB. This should be sufficient to allow for independent monitoring of intensity changes of both ranges based on amplitude. The spectra show a high level of agreement between the continuously injected and the switched case, indicating the potential of achieving similar results with even higher switching rates between the injected frequencies and thereby improving the time-resolution even further. Moreover, this approach offers a high level of in-situ tunability, as the covered spectral ranges directly depend on the modulation frequency. By fine-tuning the modulation parameters, different types of gases as well as more than two species might be accommodated simultaneously without change in device, or DC operational point.
\begin{figure*}[htbp]
\centering
\includegraphics[width=\linewidth]{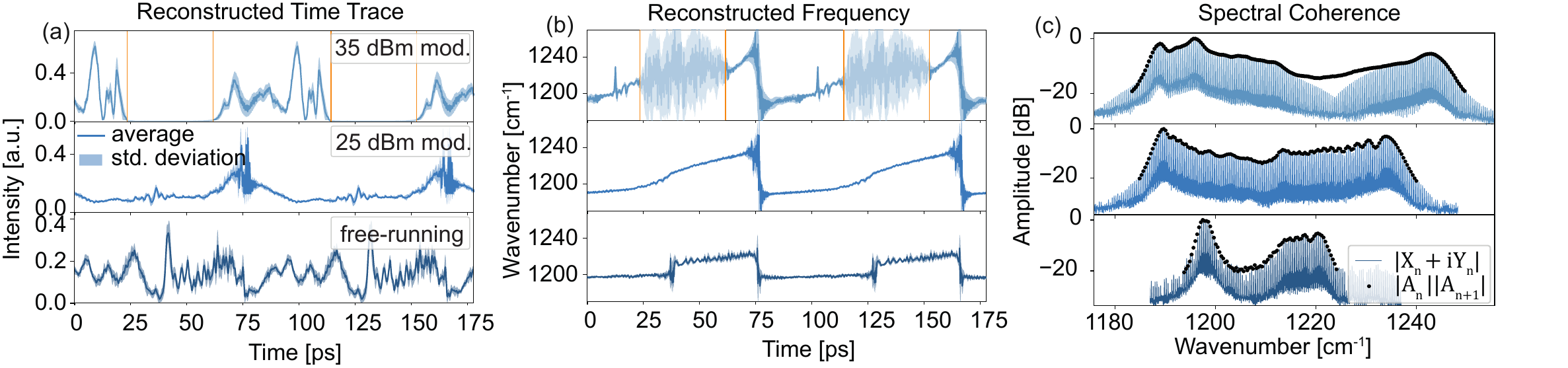}
\caption{(a) Reconstructed time-traces from the SWIFTS measurements in free-running and with 25 and 35 dBm RF-injection. The orange lines in the 35 dBm trace mark the start and end of the pulse during each round-trip. (b) Instantaneous frequency of the QCL output, as retrieved from the same SWIFTS measurements. The orange lines coincide with the ones in (a). (c) The corresponding SWIFTS spectra, the Spectrum Products are added as black dots for comparison.}
\label{fig:switching}
\end{figure*}

The effect of high-power RF-injection on the temporal shape of the QCL output can be evaluated by looking at the reconstructed time-traces from the SWIFTS measurements. In \textbf{Figure \ref{fig:switching}} a) the averaged reconstructed time-traces from SWIFTS measurements are shown at free-running operation as well as under 25 and 35 dBm injection at 11.090 GHz and 11.088 GHz, respectively. As observed in the work of Kapsalidis et al.,\textsuperscript{\cite{Kapsalidis2021Mid-infraredGeometry}} the reconstructed time-trace from the SWIFTS data changes from a quasi-continuous waveform typical for free-running QCLs to having a significant amplitude modulation at 25 dBm injection. Upon further increasing the modulation power to 35 dBm the modulation makes way for pulsed operation. The pulse-length is roughly on the order of half the round-trip time of the device. This is also apparent in the frequency traces in Figure \ref{fig:switching} b) where the uppermost trace ceases to have a well-defined frequency throughout the round-trip due to the lack of intensity in this region. Notably, there still is a linear chirp which spans the whole spectral bandwidth, where there is signal. This is analog to what has been demonstrated in ICLs\textsuperscript{\cite{Hillbrand2019PicosecondLaser}} and therefore indicates that the output could be further compressed in a suitable compressor in order to obtain shorter pulses.\textsuperscript{\cite{Singleton2019PulsesCompressor}} The comparison between SWIFTS and spectrum product in Figure \ref{fig:switching} shows a gap close to the center of the spectrum injected with 35 dBm power for this measurement series. This is a direct consequence from the sharp spectral transition at the maximum bandwidth at very high injection power. As seen in Figure \ref{fig:broadening1} c), there is a slight drop in average coherence of the spectra directly below injection of 11.088 GHz, which coincides with the frequency where the largest bandwidth occurs. Therefore, the coherence is critically dependent on the injection frequency at this point. However, the gap is at rather lower spectral power and other SWIFTS measurements with higher coherence have resulted in the same temporal shape. Therefore the impact on the disparity between reconstructed and actual output signal should be small enough to account for a weak incoherent background at most.

\begin{figure}[htbp]
\centering
\includegraphics[width=0.4\linewidth]{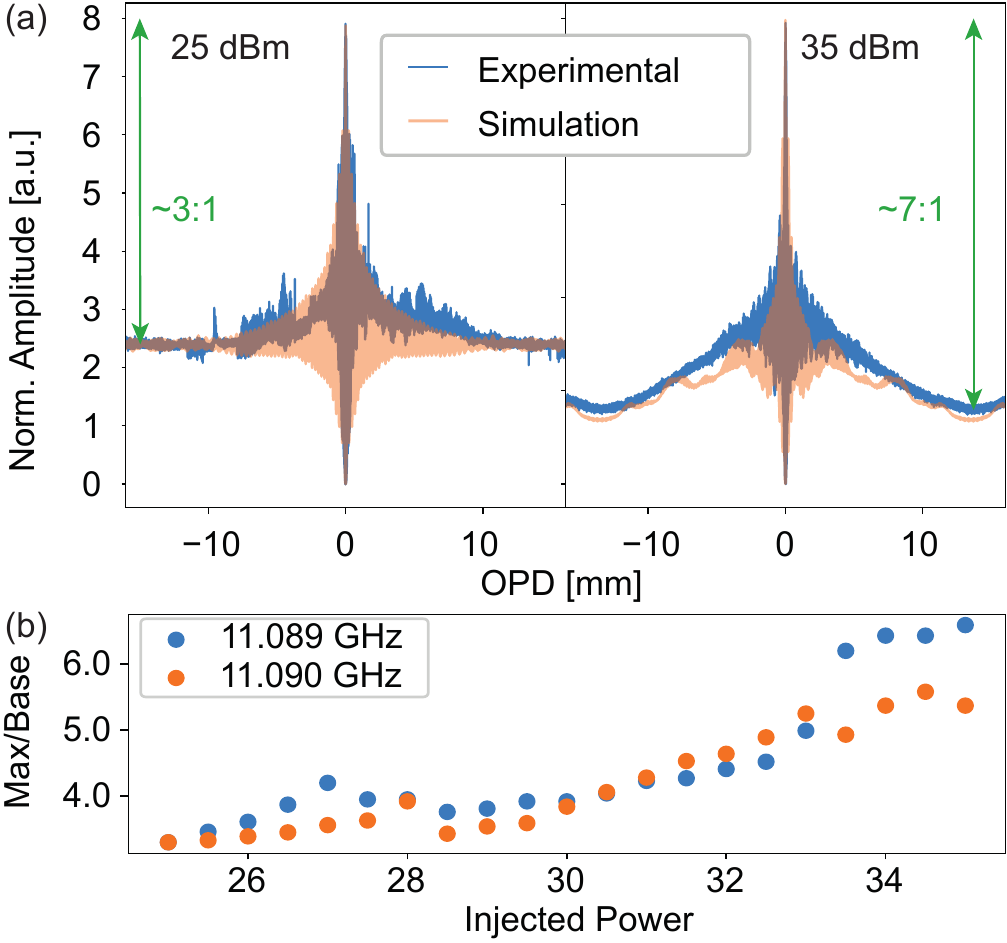}
\caption{(a) Experimental (blue) and simulated (orange) two-photon absorption data for injection levels 25 and 35 dBm. The ratio between maximum and baseline is annotated in green. (b) Maximum-baseline ratios as a function of injected RF-power at modulation frequencies 11.089 GHz (blue) and 11.090 GHz (orange).}
\label{fig:tpa}
\end{figure}

We have corroborated these results by measuring the interferometric auto-correlation (IAC) of the device under injection. Additionally, to confirm the validity of our results we simulated the IAC traces from the reconstructed SWIFTS data. The results for modulation at 25 and 35 dBm are shown in Figure \ref{fig:tpa} a). In both, simulation and experiment there is a clear difference between the strongly modulated (25 dBm) and the pulsed signal (35 dBm). The shapes of experiment and simulation agree well for both modulation powers, which further reinforces the evidence for the pulsed waveform retrieved by SWIFTS. The linear chirp during the pulses is manifest in the modulation depth of the baseline of the IAC at 35 dBm injection. It needs to be noted that in order to achieve the high incident power required for two-photon detection, any attenuating optics were removed from the beam-path, thus increasing the amount of optical feedback on the QCL. Thus, a discrepancy between experimental and simulated data from previous measurements is expected. Therefore it is worth noting that the IAC data at 35 dBm contains significantly less noise than the one at 25 dBm, indicating an increased robustness to feedback at 35 dBm. In b), the injection-power dependent maximum-to-baseline ratio is plotted as a function of injected power. The kinks in the data indicate a non-trivial relationship between injected power and degree of pulsing. For injection frequencies further from the "sweet spot" the maximum-to-baseline ratio is comparatively lower at high injection powers, which indicates that the pulse-characteristics of the output are weaker.

\section{Conclusion}
In this work we could demonstrate control over the spectral and temporal properties of a QCL frequency comb using electronic RF-injection. We could achieve consistent coherent spectral broadening throughout the device's dynamical range, which amounts to a frequency comb with an almost constant spectral distribution of the modes but a repetition rate that is tunable by up to 80 MHz. Moreover we could show rapid switching between two distinct spectral ranges by switching between injection frequencies at a switching rate limited by the capabilities of our RF-generator. In the time-domain we could demonstrate the ability to transition from quasi-CW to pulsed output, verified by IAC measurements, by ramping up the injection power.

These results show promise for improving spectroscopic applications based on QCLs, especially dual-comb spectroscopy or monolithic time-resolved monitoring of multiple gas species. Improved RF-coupling for a stronger response as well as a broadened gain bandwidth are steps that could be taken to further improve these devices. Moreover, the potential application in pulse generation  after recompression opens up the pathway to QCL-based non-linear optics, such as supercontinuum generation.

\section{Experimental Section}
\begin{figure}[htbp]
\centering
\includegraphics[width=0.5\linewidth]{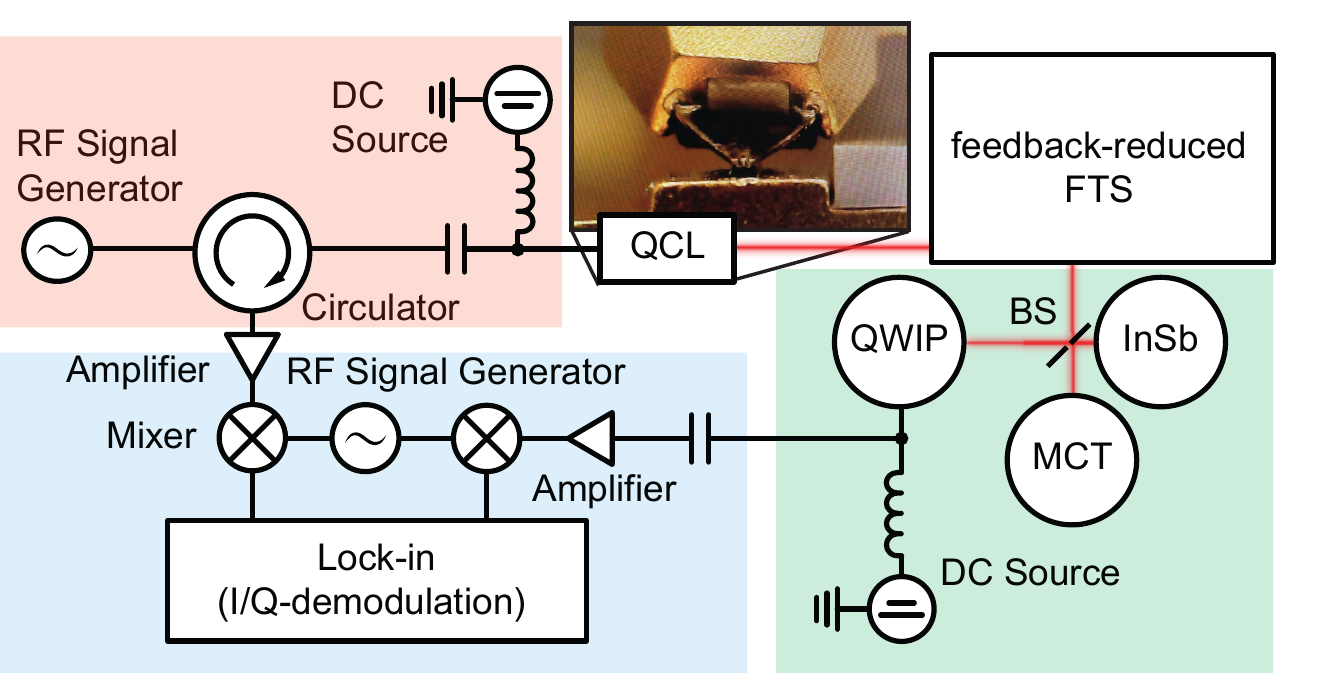}
\caption{Schematic of the experimental setup. The QCL is electrically driven through a high power DC and RF-probe (inset). An RF-circulator is connected to the circuit via a bias-tee, which allows for simultaneous injection and read-out of current modulation in the laser bias. The optical output coming from the QCL is fed into an FTS. The DC spectrum is recorded using a MCT detector. A high-speed QWIP is used to acquire the SWIFTS quadrature spectra. Further, an InSb detector with 5 µm cut-off wavelength, used as a two-photon detector, allows us to record the second-order autocorrelation of the QCL.}
\label{fig:setup}
\end{figure}

To assess the spectral and temporal response of the device, the optical output was fed through a Fourier-transform spectrometer (FTS) and onto an MCT detector. In addition, either a fast QWIP was used when performing shifted wave interference Fourier transform spectroscopy (SWIFTS), or an InSb detector with a cut-off frequency at 5 \micro m as a two-photon detector\textsuperscript{\cite{Boiko2017Mid-infraredDetectors}} in the case of interferometric autocorrelation (IAC) measurements. The measurement setup is shown in \textbf{Figure \ref{fig:setup}}. All RF-powers indicated in this paper refer to the nominal output level of our RF-synthesizer after a 35 dBm amplifier. 

\medskip
\textbf{Acknowledgements} \par
The authors would like to thank the BRIDGE program by the Swiss National Science Foundation and Innosuisse under the Project CombTrace (No.20B2-1\_176584/1) and the Qombs Project by the European Union's Horizon 2020 research and innovation program (No. 820419) for their financial support. They would also like to thank Philipp T\"aschler for proofreading of the paper and fruitful discussions.

\medskip
\textbf{Conflict of Interest} \par
The authors declare no conﬂict of interest.

\medskip

\bibliographystyle{MSP}
\bibliography{references.bib}

\end{document}